# Message in a Bottle – An Update to the Golden Record

Part I: Objectives and Key Content of the Message

Jonathan H. Jiang[1], Anamaria Berea[2], Heather Bowden[3], Prithwis Das[4], Kristen A. Fahy[1], Joseph Ginsberg[5], Robert Jew[6], Xiaoming Jiang[7], Arik Kershenbaum[8], David Kipping[9], Graham Lau[10], Karen Lewis[11], C. Isabel Nunez Lendo[12], Philip E. Rosen[13], Nick Searra[14], Stuart F. Taylor[15], John Traphagan[16]

[1.] Jet Propulsion Laboratory, California Institute of Technology, Pasadena, CA, USA
[2.] Computational and Data Sciences Department, George Mason University, VA, USA
[3.] Los Alamos National Laboratory, NM, USA
[4.] Vivekananda Mission High School, WB, India
[5.] Yeshiva University, New York, NY, USA
[6.] All Earth Citizens Foundation, Irvine, CA, USA
[7.] School of Physics and Technology, Wuhan University, Wuhan 430072, China
[8.] Department of Zoology, University of Cambridge, UK
[9.] Department of Astronomy, Columbia University, NY, USA
[10.] Blue Marble Space Institute of Science, Boulder, CO, USA
[11.] Department of Philosophy, Barnard College, Columbia University, New York, NY, USA
[12.] Climate Change Cluster, University of Technology Sydney, Ultimo, Australia
[13.] Independent Researcher, Vancouver, WA, USA
[14.] Interstellar Foundation, Johannesburg, Gauteng, South Africa
[15.] The SETI Institute, Mountain View, CA, USA
[16.] Interplanetary Initiative, Arizona State University, Tempe, AZ, USA

Corresponding: Jonathan.H.Jiang@jpl.nasa.gov

## Abstract

In the first part of this series, we delve into the foundational aspects of "Message in a Bottle" (henceforth referred to as MIAB). This study builds upon the legacy of the Voyager Golden Records, launched aboard Voyager 1 and 2 in 1977, which aimed to communicate with intelligent species beyond our world. These records not only offer a snapshot of Earth and human civilization but also represent our desire to establish contact with advanced alien civilizations. Given the absence of mutually understood signs, symbols, and semiotic conventions, MIAB, like its predecessor, uses scientific methods to design an innovative means of communication that encapsulates the story of humanity. Our goal is to share our collective knowledge, emotions, innovations, and aspirations in a way that provides a universal, yet contextually relevant, understanding of human society, the evolution of life on Earth, and our hopes and concerns for the future. Through this time and space traveling capsule, we also strive to inspire and unify current and future generations to celebrate and safeguard our shared human experience.

## 1. Introduction

As we begin to contemplate the thriving legacy of humanity in the context of potential interactions with extra-terrestrial intelligences, our future descendants or even potential alien archaeologists, a central question arises — *How should we consciously influence the portrayal of human history, ensuring its comprehensibility and significance in the event of our civilization's disappearance?* In this paper and subsequent studies, the use of "our" encompass all of humanity, transcending divisions or distinctions among human beings. The Voyager Golden

Records, an early effort at extraterrestrial communication as well as aiming to send details about Earth and its inhabitants deep into space for posterity, aspired to foster contact with evolved extraterrestrial beings. Convincing evidence of such beings, potentially residing within our galaxy and/or beyond, although widely speculated, continues to be elusive. Likewise, whether they have any past or present knowledge of the presence of life and intelligence on Earth remains an enigma.

The contemplation of communicating with extraterrestrial intelligences is by no means a recent one. The Rosetta Project, for instance, spearheaded by the Long Now Foundation, embarked on a mission to create a durable archive of 1,500 human languages intended to last for over 10,000 years [*Rose*, 2002]. Although its primary objective was to preserve languages on the brink of extinction, it also underscored the importance of language and symbols for long-lived communication, which can be extrapolated to extraterrestrial dialogues. The usage of a mathematical or universal language to communicate with extraterrestrial intelligence has been widely debated in scientific circles. Some argue that mathematics, as a universal language, holds the key to transcend interstellar barriers and cultural nuances [*DeVito*, 1990]. Freudenthal, in his book "Lincos: Design of a Language for Cosmic Intercourse" (1960), introduced "Lincos," a mathematical language specifically designed for communication with extraterrestrial intelligences [*Freudenthal*, 1960]. Mathematical languages' inherent universality, consistency, and the fact that they are not tethered to any specific earthly culture make them a promising candidate for such ambitious endeavors.

Building on this foundation, our “Message in a Bottle” (MIAB) embarks on a renewed journey to explore the age-old question posed by Roman philosopher Lucretius: Are we solitary in this boundless cosmic sea? MIAB, envisioned to traverse interstellar realms aboard an uncrewed spacecraft, holds hope that a technologically advanced civilization might one day intercept this beacon from Earth. A progeny of the Golden Records from the Voyager missions, MIAB is a modernized and meticulously articulated refinement of its forerunner.

The Arecibo Message of 1974 stands as a pivotal event in the pursuit of interstellar communication, utilizing binary arithmetic to depict human DNA, our Solar System, and human figures [*Sagan et al.*, 1978]. However, advancements in technology and research have paved the way for an updated message, as proposed by *Jiang et al.* [2022]. This contemporary message harnesses the power of modern telescopes and arrays, building upon the legacy of the Arecibo Message by incorporating an enhanced array of information, including our Solar System’s position in the Milky Way and detailed depictions of Earth and human life.

This manuscript marks the beginning of a series, outlining potential naming conventions, goals, format, and key content propositions for broader, more encompassing messages that encapsulate both Earth's and humanity's chronicles for an alien audience. While the odds of an extraterrestrial entity receiving our communiqué are slim, the replicated version safeguarded on Earth will offer invaluable insights about our civilization and its legacy to succeeding generations or, potentially, to intelligent species that might one day visit our planet. Through this initiative, our intention extends beyond just establishing contact with distant civilizations. We also aspire to establish an enduring image of humanity's essence, potentials, and achievements, both within the vastness of space and here on our home planet.

## 2. Purpose and Objective

There are two purposes for this paper and three objects for the MIAB Program. The first purpose is to propose and publicize the effort to design a message that tells the story of humanity as a way to initiate communications with an extraterrestrial intelligence or a future race of intelligent beings on Earth. Through our series of papers and promotional events, we aspire to raise interest and solicit participation, input, and feedback from citizens across the globe to generate a diverse and comprehensive view of human civilization as it exists today. In doing so, we sought to accomplish the second and equally important purpose of this paper: to unite humanity under a common goal and remind people that we are one species with a common history and will share a common future depending on how we shape the planet. The paper will inspire people to think about and provide ideas to express our origin, evolution, and accomplishments as well as bring more emphasis to science, technology, engineering, arts and math ("STEAM") education.

The Message in a Bottle initiative aims to accomplish three objectives simultaneously. Imagine if we received a message from an alien intelligence detailing themselves and their way of life. Such a confirmation of the existence of self-aware living beings in the cosmos has the potential to radically change humanity's perspective about our place in the Universe and influence the course of human history. Extraterrestrials may be pondering the same question and if they receive a message from us, it could be similarly transformative. First and foremost, the objective of MIAB is to communicate the story of humanity to extraterrestrial intelligence. If this is the first contact, the message will announce our existence and share our biology, culture, knowledge, and accomplishments as a way to promote understanding and further interactions. With a clear objective of diplomacy, we can leverage diverse and collaborative efforts across the globe to create messages that can sufficiently represent our many-faceted civilization.

To properly design the structure and content of this new message, we must imagine our audience in broad terms as we develop a clear purpose and objective. However, there are limits to this because it is difficult to imagine a type of being entirely unlike ourselves. It is entirely possible that concepts such as "civilization" may not apply in a meaningful way to an alien intelligence, but in order to proceed, it is necessary to assume an alien intelligence that will be in some way like us and can make sense out of our attempts at communication. In one scenario, our spacecraft is intercepted by a highly advanced space-faring civilization, Type II[2] or greater on the Kardashev Scale, which, despite the inherent limitations, provides one way to imagine intelligent alien life [*Kardashev,*1964]. From a logical standpoint, three potential outcomes emerge: First, they may disregard us due to the perceived cost and effort, possibly viewing humanity as too underdeveloped for meaningful interaction, akin to an ant seeking the attention of a human; second, contact and interaction occur between two species; third, they may gain knowledge about our civilization, but further interaction may be precluded by the possible extinction of humanity. In all plausible scenarios, the message plays a critical role in influencing reaction and inviting a response which can lead to greater possibilities. In order to facilitate understanding, any message should extend beyond facts and pictorial representations, opening a window into our lives and achievements, our talents and passions, and our aspirations. Nevertheless, it is crucial to show humans as an intellectual, emotional, caring species worthy of interacting with regardless of the distance, time, and energy required initiating a response. To capture their interest and imagination even in a future where humans are extinct, we should strive to make the message as engaging and informative as humanly possible.

In an alternative scenario, the spacecraft reaches a pre-Type I civilization [1] who are not yet space-faring. Our message must arrive intact on the surface of their planet like a "black box" recorder found after an airplane crash. Accordingly, we envision the structure and content of the message to be organized in two tiers, accommodating intelligible depictions for aliens with different levels of sophistication. Tier 1 content, comprising of simple illustrations of fundamental concepts and information about humanity and our environment, would a medium of transmission not requiring power or specific devices to sense, thus accessible to a pre-industrialized society. Tier 2 content adds to the degree of complexity with the inclusion of digitized information intended for alien civilizations that are closer to Type I, possessing the ability to view the content and learn about life on Earth in a more detailed and meaningful way. Even if the intended audience is not technologically capable of interacting with the digital content at the time of the encounter, they may eventually learn more about us as their civilization develops the required potential.

As the spacecraft travels through the vastness of the cosmos, with even the nearest star being 4.25 lightyears away, the most likely outcome is that it will wander endlessly without ever making contact. Alternatively, it is plausible that humanity stands alone in the vast cosmos, devoid of any other intelligent extraterrestrial entities to encounter our message. In this sobering case, the audience of the message would not be extraterrestrials but rather future generations of humans. This aligns with the message's secondary objective of fostering global unity, cooperation, a shared sense of identity, and mutual understanding among all of humanity. Through the process of designing the message that communicates what it means to be human, people will have the opportunity to reflect on and celebrate our character, culture, knowledge, environment, and accomplishments. It would stand as a demonstration to our bustling civilization of so far eight billion individuals, as well as the generations to come, of how connected we are as one species, sharing a common origin, home, and purpose. While parted in so many ways, we are also united as members of the same species across our world with a shared feeling of pride in what we have accomplished and how far we have reached into an otherwise seemingly quiet cosmos.

The third objective will be to create and preserve a comprehensive record about humanity and Earth in videos, images, and sounds as a kind of space and time capsule of human civilization and the Earth prior to the 2030s. With one version sent into deep space and another version archived on Earth, humans or other intelligent lifeforms in the distant future may find valuable historical relics. Even if Earth ceases to exist in the future, we can at least take some measure of comfort that a record of humanity survives in deep space.

With a multitude of purposes and diverse objectives, a clear strategy is required for effectively communicating the intended messages to various audiences simultaneously. In the nearly half century since Voyager, and as our robotic emissaries begin to cross the threshold into interstellar space, our society and technology has advanced substantially, enabling the creation of more detailed and representative messages.

## 3. Suggested Message Names

Initially, the delineation of nomenclature for the 'Message in a Bottle' warrants careful consideration within the framework of our collaborative global technological community. As the content of the MIAB is to be representative of all of human endeavors, the title carries a greater significance primarily due to the inclusion of content significantly sourced in part from the public. A dryly descriptive name, however dispassionately correct, may not best achieve the goal of inspiring broader wonder and engagement. A best first impression can well be the motivator towards inspiring worldwide interest and support.

A short list of possible names for the MIAB is suggested below, with their respective rationales explained in Appendix A.

1. *Earth and Our Nature (EON) Record & Archive.*
2. *Earth's Record & Archive.*
3. *Interstellar Wanderer.*
4. *Humans on Earth's Record & Archive,*
5. *大 Record.*

## 4. Suggested Content of the MIAB

The idea of *Time Capsules* has not only been an emerging realization of a more rational and scientific society, but its essence has been anticipated for millennia. In pursuit of this temptation to communicate with the future, virtually every ancient civilization attempted to create their record and in the best possible cases, these records allow us to paint a detailed picture of our own past, often giving us a *perspective of the significance of our own actions at this moment in the long journey of our species* [*Boyle*, 2017]. In this section, we focus on the most important aspect of our proposed record — *content*. Taking into account the rich and carefully designed sequence of information embedded aboard the first Voyager Golden Records, we discuss possible modifications alongside proposing newer content in order to mark our socio-economic and technological advances since the launch of the Voyager probes in 1977. We also attempt to fix the imperfections (which may lead to misinterpretation of the message) that were present in the Golden Record due to the inadequacy of the necessary technologies required for their elimination and the ethnocentric nature of much of the content [*Siegel,* 2017; *Traphagan,* 2021]. In short, the content of the updated record will not only serve the purpose of more sophisticated redundancy, but will bear the timeline of human civilization from the ancient past to the latest present and possible causes of our ascension (or extinction) into the future.

The type and content of interstellar messages have been thoroughly discussed in the past by scientists in various fields involved with SETI (Search for Extraterrestrial Intelligence) or METI (Messaging Extraterrestrial Intelligence) [*Berea,* 2018]. While communication based on human-like sensory experiences of varied types (i.e., touch, smell, taste) are evolutionarily recognized and plausible on Earth in past and future species, it is unlikely that these have evolved similarly in alien species given the coevolution of their biology with their environment, and also impossible to currently predict the biology or the environment of an alien species. Most scientists agree that communication involving physical universals (e.g., the electromagnetic (EM) spectrum, therefore through image, and sound) is more likely to also be found in alien species [*Vakoch,* 2014]. For example, vision and hearing have independently evolved on Earth several times, and this type of convergent evolution can be used as a proxy for what we are more likely to see as universal features of communication in the Universe. Another type of convergent evolution and possibly universal communication types are echolocation, and through electrical signals, but these communication types, just like touch, require direct and repeated interaction with the receiver and cannot be stored, at least with our current technology, in a medium of some sort applicable for interstellar flights.

### 4.1. Key Components of the MIAB

Figure 1 illustrates the key components we consider to be included in the MIAB, while Figure 2 (in the next subsection) illustrates the logic-based guidelines to design these components, detailed in follow-on papers. To support communicating meaningfully with an intelligent alien civilization, we have laid down three assumptions for establishing relevant parameters to guide efforts in effectively developing the structure and content of the message. Starting with our primary hypothesis, let us assume that all intelligent life with an advanced civilization, both scientifically and technologically, cannot do so without sophisticated communication and collaboration, regardless of their physiology and home planet's ecosystem. Even if they develop more advanced communication, extraterrestrials will still possess some method and system of communication that resembles language, or they might possess the capability of more advanced forms of communication beyond human perception. For our second assumption, we assume the aliens possess sensory modalities to perceive and interact with the world around them, albeit their sensory organs may be very different from those of humans. Due to physical constraints in accordance with the laws of physics across the universe, there are a limited number of likely pathways that the evolutionary process could take to develop the most efficient sensory systems. Therefore, it is reasonable to assume they have an inherent ability to sense and process stimuli from EM waves, analogous to the human visual sense, and/or oscillations in pressure, comparable to our audio sense. Although the range and/or mode of interpretation of their perception of EM wavelengths might be quite different, they would fundamentally *see* the world around them and/or *hear* sounds within a certain range of auditory frequencies. For our final and most critical assumption, we consider that aliens can integrate and process information, think logically, and learn. With these three assumptions, we have a starting point for creating a message incorporating deep and complex information sufficiently representing the intricacies of humanity. While there are still many challenges to creating a universally understandable message, these assumptions lend a measure of confidence that extraterrestrial intelligence could decipher and understand the intended meaning of a carefully crafted message.

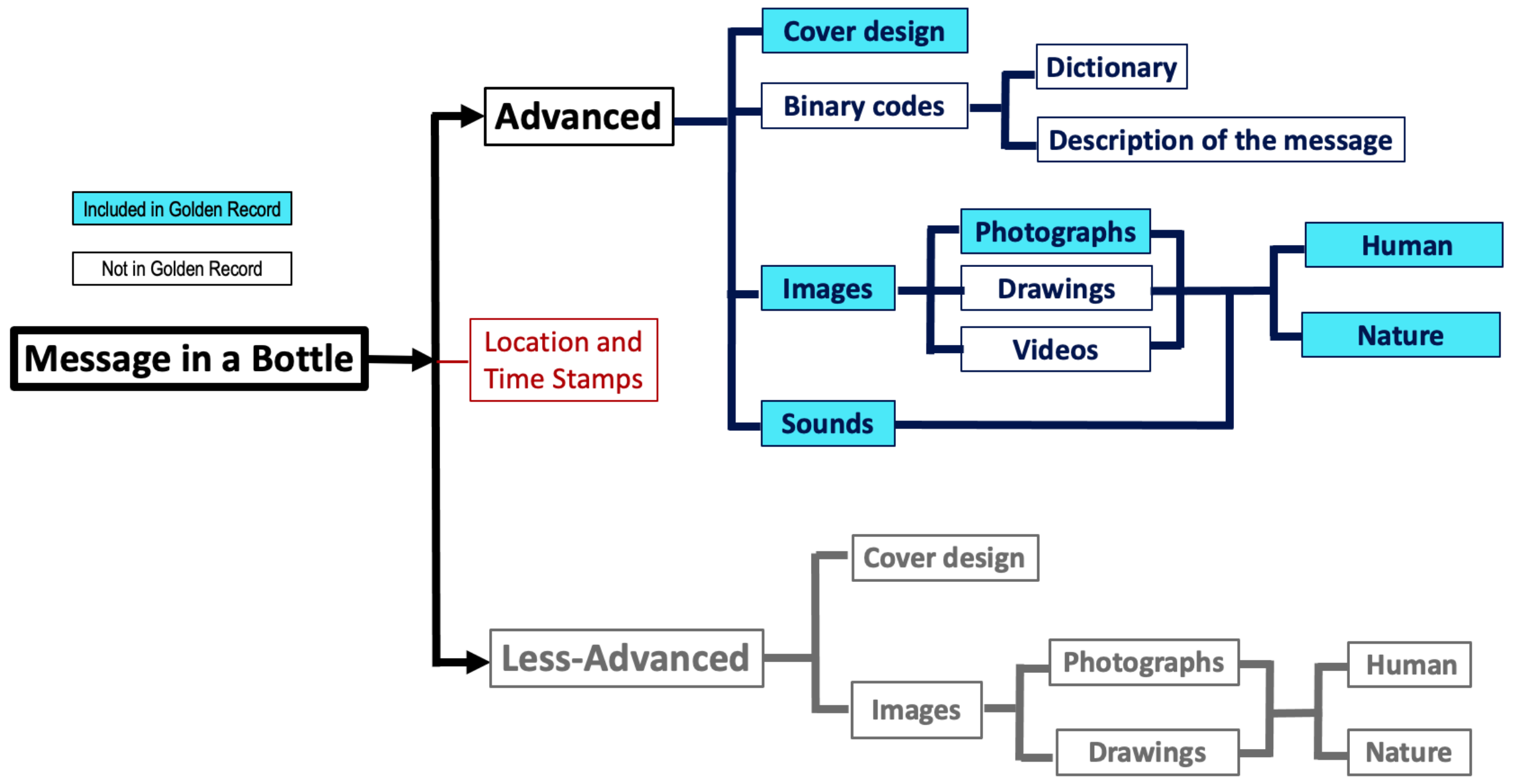

**Figure 1.** Schematic tree diagram depicting the key components of the Message in a Bottle (MIAB). The diagram provides a visual representation of the essential elements comprising the MIAB intended for potential extraterrestrial recipients. The proposed message is organized into three different sections, intended for both more and less advanced extraterrestrial civilizations. These sections encompass

scientific documentation, including Earth's spatial coordinates and a potential timestamp. Notably, the two principal sections encompass comprehensive documentation of human life and culture, with the categories of information related to these aspects highlighted in blue.

In anticipation, a twofold message is envisaged, to be encapsulated within analog and digitally electronic mediums, specifically, a *scroll* and an *advanced minicomputer*. We propose incorporating a series of images to display elementary information about humanity and Earth on the *scroll*, intended for simple direct perception (e.g., visual or tactile sensing) by extraterrestrial civilizations that are relatively less technologically advanced. For civilizations at least at the current level of human technological advancement or beyond, the record shall consist of instructions based on a symbolic representation of the procedure to reveal the digital message in the correct sequence on the *minicomputer*. Vast amounts of information would then be accessible in the form of videos, images, sounds, games and simulations, text, mathematical models, and computer code.

One challenge not easily recognized is that an interstellar message, if received by ETI, is received without any context to help interpret it. The message is not situated against any background outside of itself. By contrast, in trying to interpret the language of newly discovered people or communication among animals, one can observe the behavior of the people or animals in question. Isolated, our message will not be received against this sort of contextualizing setting. Beyond that, ETI may have no social, cultural, or historical background of humans or life on Earth against which to interpret it. Ultimately, they may exhibit fundamentally different modes of perceiving the world compared to humans. These problems extend, in a more limited way, to future species on Earth or future humans not familiar with present culture. Key to the framework for the message is to fill this lacuna, i.e., the message should provide context for itself, as much as possible. We aim to accomplish this in three ways: using videos, carefully sequencing information presentation, and linking modalities (e.g., having combined images and sounds together rather than separate).

Video (including audio) provides a way of approximating the experience of observing humanity directly. It also circumvents the issue of ETI first having to decode a symbolic system to understand the initial part of the message (though the instructions on how to play the video would have to be simply represented.) The message will be hierarchical: each layer serving as the key to unlock the next, more complex layer, guided by the principle that each should be understandable and informative. In this way, earlier layers of video provide context for later layers; for example, the concept of music would be introduced in a simple way before major musical works and the importance of music in cultures around the world are conveyed. Furthermore, video provides context beyond that of separate images and sounds. Taking again music as an example, a video of a band playing a song has a better chance of being interpreted for what it is than an image of an instrument and a disconnected sound recording of a song.

The *Introductory Section* begins a scientific conversation through establishment of a common understanding using universal laws of nature on Earth and logical operations capable of bridging the gap between our comprehensions to that of recipients. Some instances include discussion of basic mathematics, constancy of the speed of light in vacuum, and the emission spectra of commonly occurring elements. We then introduce the Earth, our Solar System and galaxy, based on fundamental concepts and facts that should be relatively straightforward to interpret by scientifically adept ETIs. Once we establish a basis for communication, it would follow to introduce translation tools with reference tables and dictionaries to create some building blocks and framework upon which more detailed communications would be rendered relatively simple. The second, *Teaching/Interactive Section*, will transition to topics about humanity and life on Earth which may be less familiar. Upon first contact, alien lifeforms would likely have little in common with humans and may not comprehend the intended purpose and meaning of the message. We therefore plan to include encoding and decoding keys for message encryption. Effective communication between two parties should be based on some commonalities and pre-established understanding regarding each other, thus the initial step would be to educate and initiate interactions with our message to develop shared experiences and a fundamental understanding. Accordingly, this section constitutes elementary information about various pertinent topics such as our emotions, social structure, culture, art, music and history. In addition to videos, sounds, text and images, we can include interactive activities and games to engage the recipients. By guiding the aliens to integrate some fundamental knowledge, in addition to glimpses of background and historical information, we can provide the foundation and context to rightly start understanding the more complex yet meaningful content that follows in the next section. Building on the second section with advanced knowledge and complex concepts, the third *Detailed Information Section* contains topics such as knowledge, technology, art, music, politics, culture, and economics, which represent the forefront of human ingenuity and creation. For economics, the second section will introduce basic concepts like trade interaction, supply and demand, and the value of money, while the third section will display the intricacies of global commodities markets, dynamic equity and debt markets, and intricate supply chains.

The taxa of knowledge that we plan to include will follow the principles developed by epistemologists and ontologists with respect to the organization and representation of knowledge as we have it today in our civilization. In computer science and information science, an ontology encompasses a representation, formal naming, and definition of the categories, properties, and relations between the concepts, data, and entities that substantiate one, many, or all domains of discourse. More simply, ontology is a way of showing the properties of a subject area and how they are related by defining a set of concepts and categories that represent the subject. The three main branches of knowledge in the tree are: "Memory"/History, "Reason"/Philosophy, and "Imagination"/Poetry, as originally devised by Francis Bacon, and which largely also follow the epistemological classifications into perception, memory, introspection, inference and testimony. Additionally, we will use the classifications of science, cultures, and examples of what is a human, a society and life itself.

The team that designed and produced the original Voyager Golden Record did so in a span of mere weeks while using 1970s technology. The effort to update that record into its new form, the *Message in a Bottle,* allows for years of methodical planning and nearly half a century of technological advancement to design and produce content more thoroughly representative of humanity. Along with leveraging these expanded resources, the intent is to also engage the citizens of the world to provide ideas for the videos, images, sounds, games, and other digital content that will convey the essence of humanity and what life is like on Earth. The digital content shall explain what it means to hope and dream, to love and care about each other, to explore and derive new knowledge, and to create art and music. The MIAB Team will solicit input, content submissions, and feedback and evaluations from people across the world. Special attention will be given to garnering information from less developed regions with rich cultural history in Africa, Asia, the Middle East, Central and South America to ensure sufficient representation across different cultures. There will be online and offline informational and educational seminars, awareness campaigns, promotional events, and contests to target groups of people from all walks of life and different age groups. This global effort will provide many perspectives and diverse ideas which illustrate the collective intelligence and experience of the human species. While all these components, though carefully engineered, may well make the message complex and thus difficult to understand, it is nonetheless worthwhile as it is imperative to capture the full array of humanity.

Harkening back to our past, but with an emphasis onto a detailed representation of the present scenario of human life and technological advancements, the contents of the newly designed message must incorporate the ideas of the Voyager Golden Records while including newer facts of the present. All of which, taken as a whole, shall be critical for a complete narration of our story – with whomever it may come to rest in the vast, seemingly endless ocean of the cosmos. In this context, the key components should include two broad categories based on the form in which the content would be encoded: imagery and sounds from Earth. Audio will not only include sounds from our natural surroundings, but also incorporate music and a myriad of others commonly encountered in our lives along with impressions of the modern and technologically advanced societal structure of which we are all a part. It should be noted, however, some images and their associated sounds should be connected, preferably by video, as de-contextualized sounds of Earth would be almost impossible to decipher. Considering a generally similar approach to that opted for in the content selection of the Voyager Golden Records, we shall as well suggest modifications while also including the most relevant content spanning the two generations which have elapsed. We have further classified the imagery content into two subsections that would address the three most important questions the recipients might be concerned with: the origin of the record, the senders of the record, and most importantly, a detailed overview of our nature.

### 4.2. Methodology for Key Components Design

The logic flow used in the overall construction of the MIAB content is illustrated in Figure 2. Essential in this endeavor are clearly stated controlling concepts serving as guiding principles, forming the conceptual foundation upon which the MIAB will be built. First, the location and timing of the MIAB's origin is articulated in objectively scientific terms, placing the message in context within the space-time scale it is expected to traverse. Leveraging the methods of prior messages to ETIs afford an efficient means of accomplishing this portion. Next, the MIAB will be engineered as an emissary of Earth, encompassing the diversity of humanity, our planet including non-human life, and aspirations for contact. Finally, in weighing the multitude of information to include, we employ as a sounding board what we believe an emerging spacefaring civilization such as ours would want to receive in a message from an ETI.

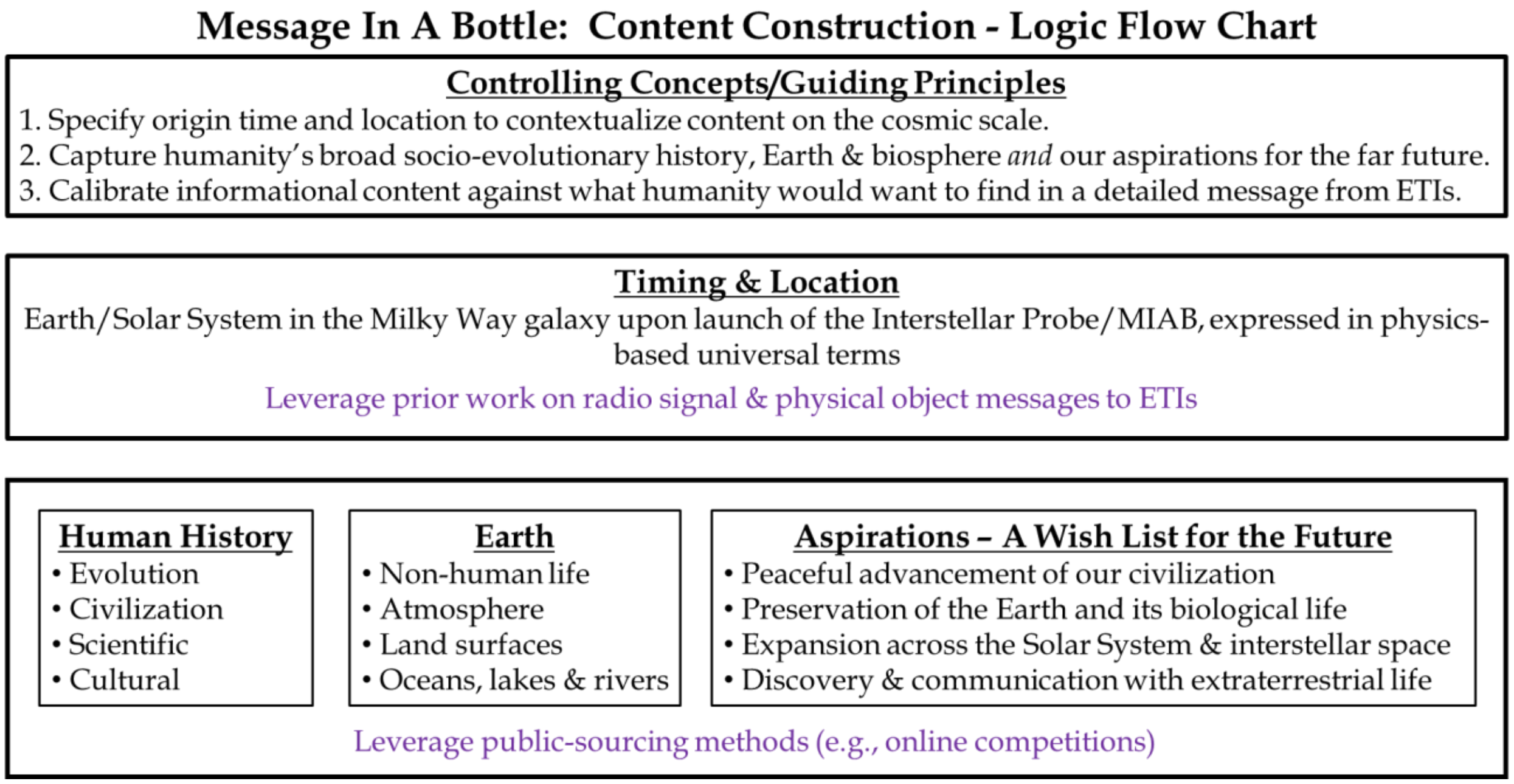

**Figure 2:** Flow chart illustrating the logical basis and methodology used in constructing the MIAB's key components.

The bulk of the MIAB, which using present day technology could at least be derived in part from public input, will consist of a history of its creator species, descriptions of humanity's home world, and examples of human hopes for the future. In describing ourselves, we logically begin with the long path evolution has taken us, starting from the simplest of lifeforms, followed by the emergence of civilization in various forms across the last several thousand years. Enclosed within this description are the glimpses of our scientific achievements such as nuclear fission and space exploration, along with examples of wide array of cultures and knowledge which comprises the complex human tapestry. Having introduced ourselves, the physical Earth is then depicted in detail accompanied by an overview of the biosphere across the range of environments life has come to fill.

In completing the MIAB, our aim is to look ahead to the far future. Our message can also be thought of as a testimonial, sent on behalf of our world and its life. To fulfill this role, an encapsulation of who we are in the present must also extend to visions of what we might become – in short, examples of human aspirations. Though assembled in a present whose troubles are all too apparent, the MIAB should speak to a future of peaceful, sustainable, and diverse development of humanity. In concert, such aspirations would see Earth and its biosphere preserved and protected even as humans find their way to other worlds in the Solar System, and eventually worlds around other stars, our new homes. Returning to the most basic purpose, we conclude with an invitation to any ETI who may encounter the MIAB to reply with their own story. In summary, the MIAB's key components, collected and constructed via the logic flow described, will offer a comprehensive and eloquent statement worthy of humanity now and into the far future.

### 4.3. Dictionaries

Dictionaries serve as one of the most efficient means of conveying large quantities of data due their immense potential for the symbolic and careful inclusion of rich and interconnected pieces of information through the utilization of concise imagery. Besides acting as a representative of the hierarchical classification of imagery chronicled with an increasing level of complexity, this as exemplified for MIAB in Figure 3, dictionaries exhibit their unique aptitude in the form of index tables, which reshape this genre into a universally compatible set of information. Often in conjunction with videography, the transfiguring notion of visualizing the world in unison with auditory effects present an exceptional impact on the proposed next-generation of advanced content for the updated record. Dictionaries were a major constituent of the Voyager Golden Records and their forerunners, the Pioneer plaques, were themselves a type of dictionary, although with relatively lesser content.

For MIAB, we will utilize dictionaries as a means of not only detailing a set of answers, but also to encompass necessary prior universal knowledge that would indeed facilitate the understanding of our conventions and the art of scientific communication from the perspective of the intended recipients. Therefore, the proposed dictionaries require careful design methodology with an absolute priority on logic, a circumstantial discussion of which shall be laid out in further studies in this series. As a starting point, the projection of these distinctive documents might be two-fold, with a former class incorporating a detailed overview about the basic understanding of mathematics and physics (thus, also serving as a tool to unlock the rest of the quantitative information present in the record) and a latter class conveying the more complex yet pivotal information about our origins, i.e., the Solar System and our Earth as well as the major components that have made life possible on planet Earth, thereby signifying our home world as a planet where intelligent life is prevalent.

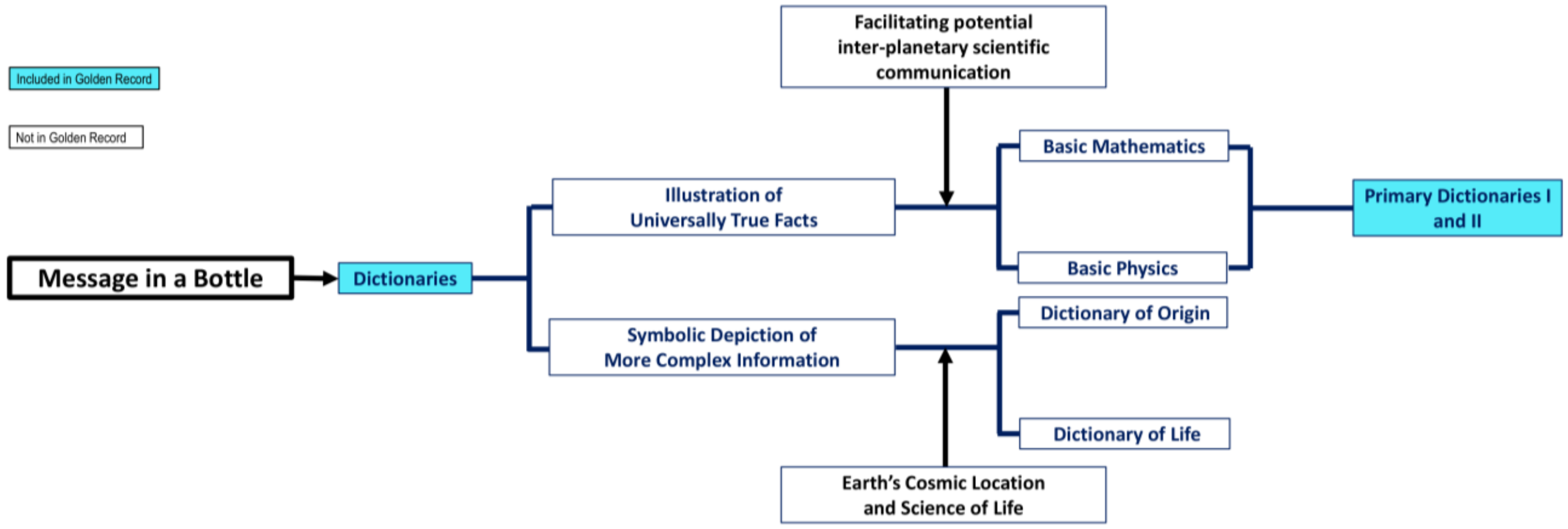

**Figure 3:** Diagrammatic representation of MIAB's Structural Design primarily based on Dictionaries. The illustration portrays the construct, predominantly reliant on dictionaries. The content delineates two distinct sectors, encompassing fundamental elements of universal science and the depiction of critical scientific information about life on Earth. Three interconnected dictionaries are proposed, with the initial two being adaptations of those encoded within the Golden Records, representing fundamental principles of mathematics and physics.

## 4.4 Our Location and Time Stamps

### 4.4.1 Location Stamp

One of the monumental aspects critical towards the careful design and construction of the proposed message is the inclusion of a logical representation indicative of our place in the cosmos. Assuming that the recipients are at least at the level of technological and scientific advancement as we are now, the need for schematic representation puts forth a myriad of diverse and wide-ranging options in terms of interpretable astronomical details. In other words, the design of an effective *galactic positioning system* (*GPS*) would take us one step closer towards the establishment of a generalized location map for potential alien races to discover humanity and in-turn, our home on this Pale Blue Dot.

For MIAB, we propose implementation of a GPS via application of globular clusters (GCs) as the founding principle and millisecond pulsars acting as secondary indicators in the creation of the location map. An exemplary portion of this methodology is illustrated in Figure 4. This new map is based on a systemic inclusion of GCs' properties, such as luminosities, metallicities, and other observed factors. A detailed explanation of the proposed location map is presented in Appendix B.

### 4.4.2 Time Stamp

In general, due to the substructure evolution in the galaxy, it is critical to specify the design and launch time of the proposed location map. Otherwise, though future life may decode the map successfully, they would not necessarily realize the timeline of human existence and as a consequence, will not be able to manifest the galactic scenario at a specific time in the past. Fortunately, proper motion makes the map itself time-dependent, thus making the GC map itself encode time information in the relative location of globular clusters. Combining Gaia's long-period precise location measurements and ESO or Keck's radial velocity data, the three-dimensional proper motion velocities of 164 globular clusters are available [*Baumgardt et al.*, 2023].

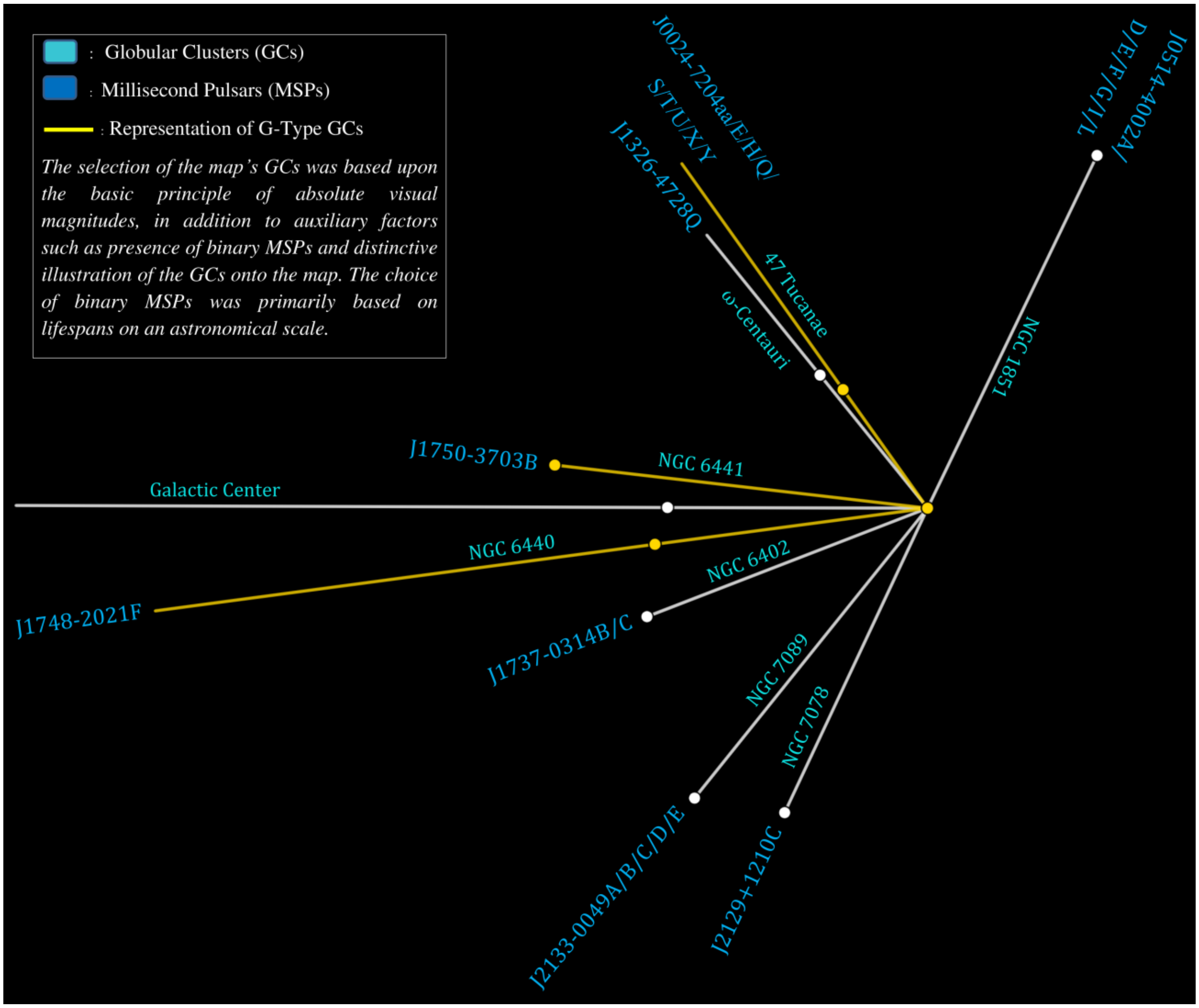

**Figure 4:** The proposed GC-Based Location Map for MIAB. This schematic representation depicts a celestial location map that relies on the longitudinal distribution of globular clusters (GCs) as reference points with reference to the galactic center, with each GC indicated by its respective Millisecond Pulsar (MSP) marker(s). GCs, known for their stability, serve as crucial cosmic landmarks for interstellar navigation and communication, with the inclusion of plausible MSPs enhancing precision in celestial positioning.

As illustrated in the left-panel of Figure 5, the proper motion velocities are distributed in the range of 0 to 400 km/s, with a large majority being over 100 km/s. According to the GC map's spatial resolution of 1 ly, this can be converted to time resolution easily. Curves in Figure 5's right-panel plot the time resolution of the GC map at different proper motion velocities and different spatial resolutions. As to the 1 ly space resolution GC map, its time resolution can reach 1000 years easily, once the GC's proper motion velocity exceeds 300 km/s. If we lower the spatial resolution to 10 ly to reduce the message size, the 10,000-year resolution also can be achieved.

However, the time stamp precision not only depends on spatial resolution but on location precision as well. Three-dimensional location measurements of objects are still limited by observational capability, especially the distance measurement. For example, for a 0.5 kpc distance error coupled with 500 km/s proper motion, the corresponding time is about $10^6$ years, much higher than the time resolution as we mention above. If we have a more precise distance, such as an error lower than 0.1 kpc, with the 300 km/s proper motion the time error will decrease to $3.3 \times 10^5$ years – a substantial improvement. Therefore, we suggest implication of location mapping as a medium of time stamping in itself. Further developments in observational technology, enabling the locations of GCs to be known with greater precision, would foster even higher time resolution measurements.

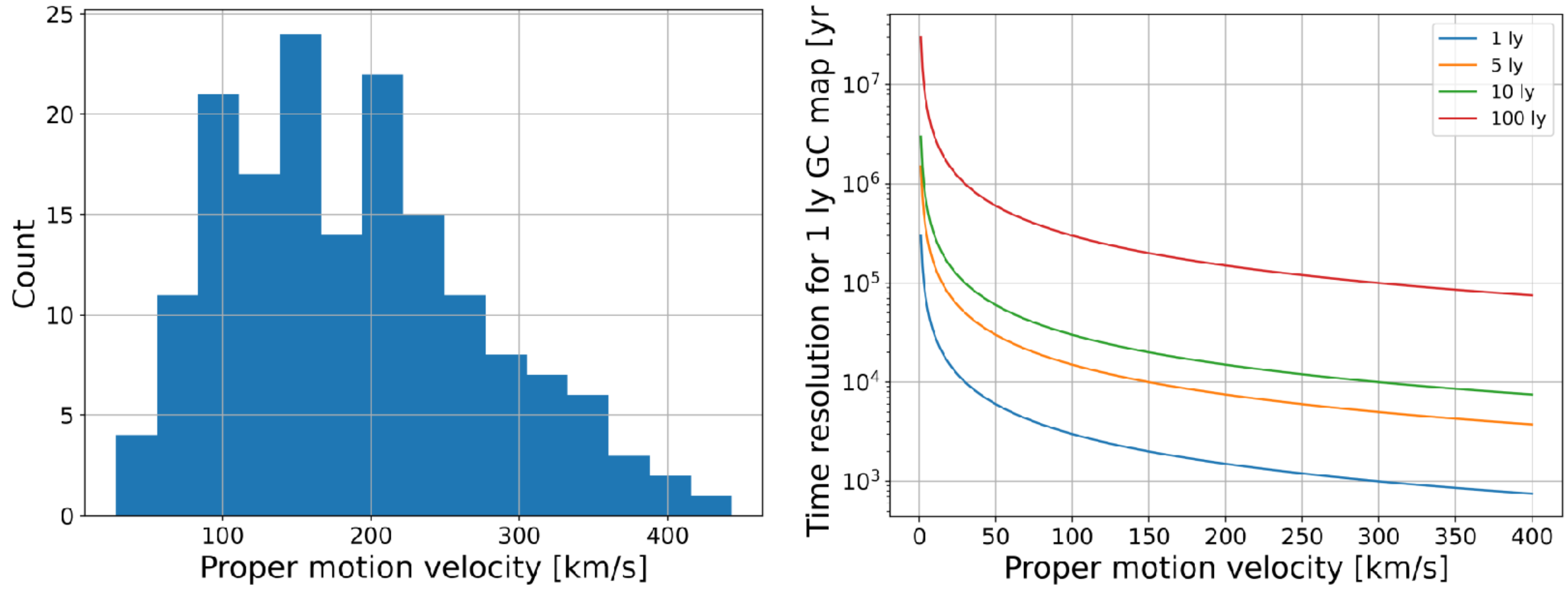

**Figure 5:** Left panel: Distribution of 164 GCs' proper motion velocity with orbital integrations were done in the *Irrgang et al. (2013)* galactic model; Right panel: Time resolution of the GC map at different proper motion velocities for four spatial resolution cases.

## 5. Conclusion and Discussion

With the acceleration in our societal evolution towards a Type I civilization[1] and beyond, we have come to the point of questioning the long-term continuation of our own existence. The relentless utilization of natural resources, with a still developing grasp towards renewal technology, is evident in our continued reliance on fossil fuels *which puts the global economy and energy security at the mercy of geopolitical shocks and crises* (as stated by the Secretary General of the United Nations at the 2021 United Nations Climate Change Conference.) Hedging against scenarios of this and other potential Great Filter events [*Jiang et al.*, 2023] in the near or distant future that might threaten the survival of our race, carefully designed and well-preserved time capsules can serve as an effective means of carrying into the future the legacy of humankind. The MIAB is one such bold approach towards representing humanity in all its complexity. The notion devises a logically universal symbolic and pictorial form of communication interpretable by intelligent extraterrestrial civilizations possessing a level of advancement at least equivalent to its senders. Starting from a foundational working design based on ideas of the creators of Voyagers' Golden Record, we strive to make the MIAB a state-of-the-art implementation of well-reasoned facts in an increasing order of complexity. To accomplish this goal, the message is communicated through discrete yet interconnected segments of information, depicted in modes of imagery, audio-visual and textual documentation.

Considering the dual scenarios of potentially less and more advanced civilizations relative to our current level of technological advancement, the record is being crafted to incorporate meaningful content interpretable by both classifications of civilizations, dictated over the two-fold structural design of the key components of the record. Besides the primary focus on implementation aspects, we have analyzed a series of possible nomenclatures which embody the core principles of the proposed record. These include communication with extra-terrestrial intelligence, an Earth-based archive and the much greater near-term objective of global STEAM educational campaigns, thus keeping the project aligned with promotion of citizen science. The methodology that we intend to apply for the careful construction and selection of the key components encompasses a three-tier logic flow. Involved is a diagrammatic representation of the central theme of the mission and proposed guiding principles. Contained within is the high-level detailing of the steps involved in efficient arrangement of message contents. In conclusion, we pose an in-depth discussion on the two major components of the record, namely, *dictionaries* and *location and timestamps*. These play the key role of illustrating the importance of the message and establishing an initial connection between the minds of the recipients and humans.

The first part of MIAB, being an introduction to the workflow of the anticipated mission, opens up the gateway to innumerable possibilities of relevant content to be discussed in subsequent papers concerning specific aspects of the curated contents. Taken in sum, the MIAB acts as the cornerstone to symbolizing humanity's presence in the vastness of the cosmos – and our desire to know of others.

**Acknowledgement:** This work was supported by the Jet Propulsion Laboratory, California Institute of Technology, under contract with NASA. We also acknowledge the supports from the George Mason University, Los Alamos National Laboratory, Vivekananda Mission High School, All Earth Citizens Foundation, Wuhan University, University of Cambridge, Columbia University, Barnard College, University of Technology Sydney, Interstellar Foundation, SETI Institute, and Waseda University.

**Data availability:** The data and software used for location and timestamps for MIAB can be downloaded from https://github.com/Prithwis-2023/MiaB-Location-and-Time-Stamps/tree/main. For additional questions regarding the data sharing, please contact the corresponding author at Jonathan.H.Jiang@jpl.nasa.gov.

**Footnotes**

[1] A Type I civilization, as postulated in the Kardashev Scale, can harness and control all available resources on its home planet, including geological, atmospheric, and oceanic phenomena.

[2] A Type II civilization, in the Kardashev Scale, can harness and control the energy of its entire star (like collecting the energy of the sun without relying on the mere collection of sunlight), extending its influence over its solar system.